\documentstyle[twoside,fleqn,espcrc2,epsf]{article}

\newcommand{\AmS}{{\protect\the\textfont2
  A\kern-.1667em\lower.5ex\hbox{M}\kern-.125emS}}
\def\mxfigura#1#2#3#4{
  \begin{figure}[hbtp]
  \vspace*{-.5cm}
    \begin{center}
      \epsfxsize=#1
      \leavevmode
      \epsffile{#2}
     \end{center}
     \vspace*{-1.25cm}
    \caption{#3}
    \vspace*{-.5cm}
    \label{#4}
  \end{figure} }

\def\abs#1{\left| #1\right|}
\def\etal{\hbox{\it et al.}}
\def\spa{\hspace*{-.65cm}}
\def\pl#1#2#3{    {\it Phys. Lett. }{\bf #1} (19#2) #3}
\def\np#1#2#3{    {\it Nucl. Phys. }{\bf #1} (19#2) #3}
\def\pr#1#2#3{    {\it Phys. Rev. }{\bf #1} (19#2) #3}
\def\prl#1#2#3{    {\it Phys. Rev. Lett. }{\bf #1} (19#2) #3}
\def\zp#1#2#3{    {\it Zeit. f\"ur Physik }{\bf #1} (19#2) #3}
\def\Re{\mathrm{Re}}
\def\Im{\mathrm{Im}}
\def\eq#1{{Eq.(\ref{#1})}}

\def\eqs#1#2{{Eqs. (\ref{#1}--\ref{#2})}}
\title{\vskip-3\baselineskip\hfill FTUV/98-87\\
\hfill IFIC/98-88\phantom{X}\\
$\tau$ weak dipole moments from azimuthal asymmetries 
       \thanks{Work supported in part  by CICYT, under Grant AEN 96-1718, 
       and by Programa de Cooperaci\'on
Cient{\'{\i}}fica con Iberoam\'erica (AECI), and by
CSIC-Uruguay.}\\
{\normalsize Talk given at the TAU'98 Workshop, 14-17 September 1998, 
Santander, Spain.}}

\author{J. Vidal\address{Departament de F{\'{\i}}sica Te\`orica,
Universitat de Val\`encia
\\ E-46100 Burjassot, Spain}$^{\rm ,}$\address{IFIC, Centre Mixt Univ. Val\`encia-CSIC
\\ E-46100 Burjassot, Spain},
J. Bernab\'eu$^{\rm a}$
and G. Gonz\'alez-Sprinberg\address{Instituto de F{\'{\i}}sica, Facultad de
Ciencias, Universidad de la Rep\'ublica\\
CP 10773, 11200 Montevideo, Uruguay}
}
       
\begin{document}
\thispagestyle{empty}
\begin{abstract}
We show that transverse and normal single-$tau$ polarization
of $tau$ pairs produced at $e^+$
$e^-$ unpolarized collisions, at the $Z$ peak, are sensitive to 
weak (magnetic and electric) dipole
moments of the $tau$. We also show how these components of the $\tau$
polarization are accessible by measuring appropriate azimuthal asymmetries
in the angular distribution of its decay
products. Sensitivities of the order of $10^{-18}$ $e\cdot cm$, for the 
weak-electric dipole
moment, and $10^{-4}$ ($10^{-3}$), for the real (imaginary) part of the
weak-magnetic dipole moment of $\tau$, may be achieved. Compatible bounds 
are also presented from spin-spin correlated asymmetries.
\end{abstract}

\maketitle

\section{INTRODUCTION}

Electron and muon dipole moments provide very precise tests of
quantum field theories.  The agreement between the predicted (first obtained 
by Schwinger \cite{sch} in first order) and the measured electron anomalous 
magnetic moment
is one of the most spectacular achievements of quantum field theory.
Electric and weak-electric dipole moments have been exhaustively
investigated to look for signals of  $CP$-violation in  both the quark and
the leptonic sector\cite{jks}. Low energy, LEP1 and 
SLC, experiments
result in an enormous variety of measurements  that 
lead, up to now, to the confirmation of the quantum corrections given 
by the  Standard Model (SM). 

The theoretical and experimental situation for the electro-magnetic ({\it i.e.}
the ones related to the $\gamma$--coupling) dipole moments of light fermions
is firmly established\cite{pdg}: experiments are sensitive to an impressive
number of decimal places and
theoretical predictions of higher orders have been
computed.
For heavy fermions ($\tau , b, t$), the magnetic  DM are much poorly measured,
and also their theoretical significance  is  more involved.  The anomalous
weak magnetic dipole
moments have been calculated for heavy fermions \cite{tau,quarks} in the
Standard Model. For $\tau$'s, weak dipole moments have  
been tested at LEP1 and SLC in recent years \cite{others} by means
of the angular distribution of the $\tau$ decay products 
acting as spin analyzers. 

In this contribution we first show how, for $e^+\, e^- \longrightarrow 
\tau^+ \tau^-$ unpolarized scattering at the $Z$-peak, the
transverse (within the collision
plane) and normal (to the collision plane)
 single $\tau$ polarizations are very sensitive to the
anomalous weak-magnetic ($a_{\tau}^w(M_Z^2)$) and weak-electric
($d_\tau^w(M_Z^2)$) dipole form factors. We construct azimuthal asymmetries,
for single $tau$ decay products, sensitive to each effective coupling in order 
to separate this signal in the search for new physics. 
Finally we present 
how some  azimuthal asymmetries coming from spin-spin correlations can 
help in the search for signals of the 
weak dipole moments.

\section{DIPOLE MOMENTS}

The most general Lorentz invariant structure describing the
interaction of a vector boson $V$  with two fermions $f\bar{f}$ can be 
written in terms of ten form factors:

\begin{eqnarray}
&&\spa
<f(p_-)\bar{f}(p_+)|J^\mu(0)|0>=\\
&&\spa e\, \bar{u}(p_-) \left[  \gamma^\mu (f_1-f_2\gamma_5)
+(i\, f_3-f_4\gamma_5)\sigma^{\mu\nu}q_\nu \right.\nonumber \\
&&\spa \left.+(f_5+i\, f_6\gamma_5)(q_-)^\mu 
+\sigma^{\nu\mu}(q_-)_\nu(f_7+i\, f_8\gamma_5)\right.\nonumber\\
&&\spa \left.+(i\, f_9+f_{10}\gamma_5)q^\mu\ \right]v(p_+)\nonumber
\label{eq:1}
\end{eqnarray}

\noindent with $q=p_++p_-$ and $q_-=p_+-p_-$. For all particles on-shell, 
these ten form factors may be reduced to the first four: $f_1$ and  $f_2$
parameterize the vector and vector-axial sector of the current; $f_3$ and
$f_4$ are proportional to the anomalous magnetic ($a_f^V$) and 
electric ($d_f^V$) dipole moment, respectively:

\begin{equation}
f_3(q^2)=\frac{a_f^V(q^2)}{2m_f},\hspace*{.5cm} f_4(q^2)=\frac{d_f^V(q^2)}{e}
\end{equation}

\noindent Dipole moment couplings can be also seen as the coefficients of the 
corresponding vector boson-fermion-fermion ($V\psi\psi$) interaction terms 
of an $U(1)$-invariant effective lagrangean 
\vspace{.25\baselineskip}
\begin{eqnarray}
{\cal L}= {\cal L}_{SM}&-& \frac{i}{2} \; d_f^V\;
\bar{\psi} \sigma^{\mu\nu} \gamma_5 \psi\; {\cal F}_{\mu\nu}\nonumber \\
 &+&\frac{1}{2}\; \frac{e a_f^V}{2 m_f} 
\bar{\psi}\sigma^{\mu\nu}  \psi {\cal F}_{\mu\nu}\label{lagrangean}
\end{eqnarray}

\noindent where ${\cal L}_{SM}$ is the tree-level Standard Model lagrangean
and

\begin{equation}
{\cal F}_{\mu\nu}=\partial_\mu V_\nu -\partial_\nu V_\mu, 
\hspace{.25cm}V=\gamma,\, Z 
\end{equation}

For on-shell photons ($V=\gamma$),  we find the usual definition for the
anomalous magnetic dipole moment (AMDM) $a^\gamma_f(q^2=0)$ and electric
dipole moment (EDM) $d^\gamma_f(q^2=0)$.  For on-shell $Z$-bosons ($V=Z$)
we {\it define}, by analogy, the anomalous weak magnetic and weak 
electric dipole moments (AWMDM and WEDM) of the fermion $f$ as the 
corresponding factors $a^w_f(q^2=M_Z^2)$ and
$d^w_f(q^2=M_Z^2)$ in \eq{lagrangean}. 

\subsection{$C$, $P$ and $T$ transformation properties\label{properties}}

As can be seen
from \eq{lagrangean}, the dipole moment
form factors  are related with chirality-flipping operators of the
theory. Under discrete $C$, $P$, and $T$ symmetries the
term with $a_f^V$ is $C(+)$, $P(+)$ and $T(+)$; the one with
$d_f^V$ transforms as $C(+)$, $P(-)$ and $T(-)$. For the $V=Z$ case, the
dipole moments may get an imaginary (absorptive) part  which, contrary to the
real part, is $T(-)$ for the A(W)MDM ($a_f$), and $T(+)$ for the (W)EDM ($d_f$). 

\subsection{Theoretical predictions}

From the theoretical point of view, as it is already well known, only the
on-shell definition of the AWMDM is electroweak gauge invariant and free of
uncertainties. The AWMDM may receive contributions from both new physics and 
electroweak radiative correction to the SM. 
\mxfigura{7.75cm}{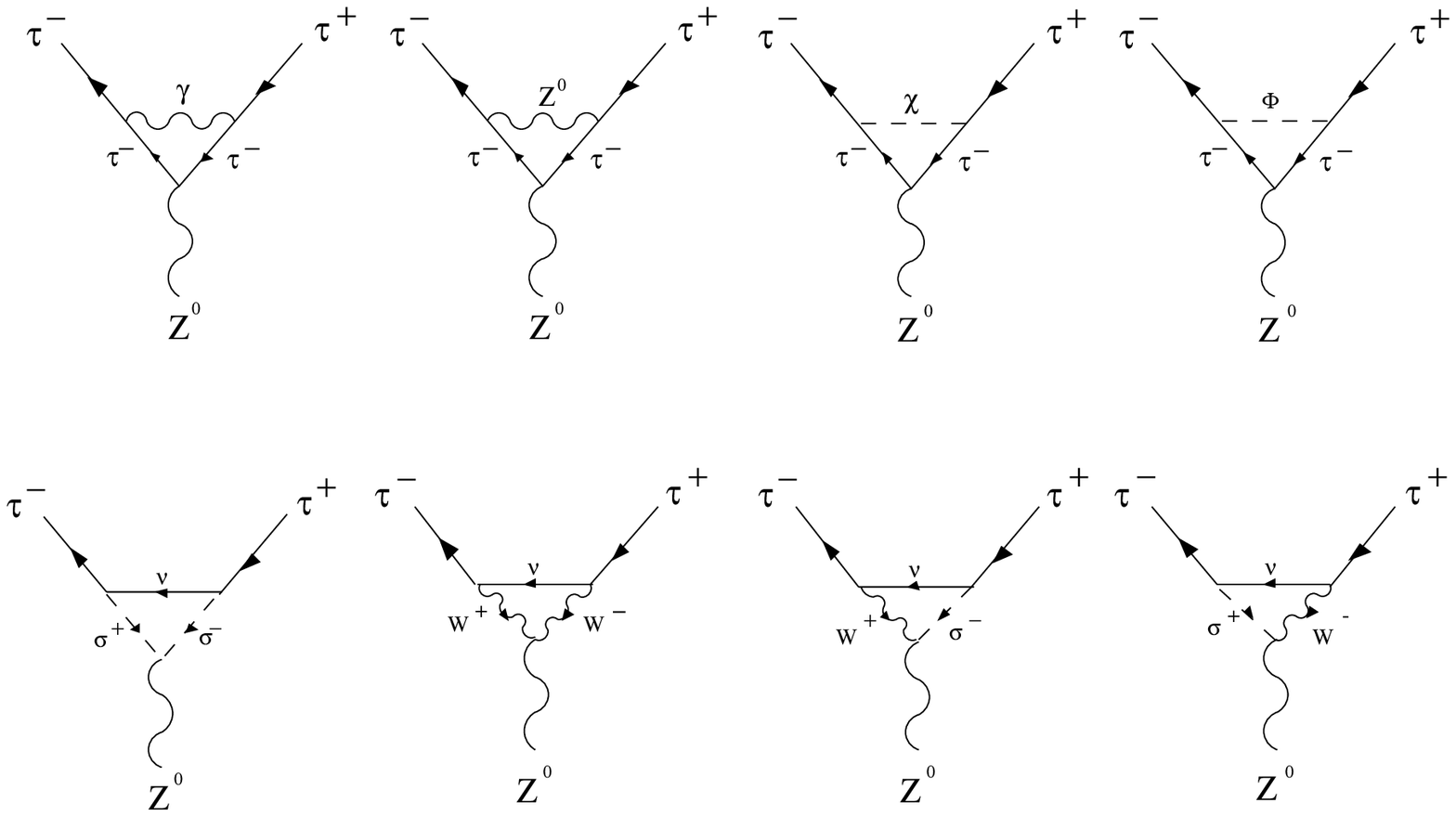}{Contributing Feynman diagrams to $a_\tau^w$ that
are also present for the AMDM (photon vertex).}{fig1}

The leading Standard Model contribution to $a^w_\tau$ has been computed
\cite{tau} in the t'Hooft-Feynman gauge, where no ambiguities in the
finite parts are present \cite{fuj}.  There are 14  diagrams to consider,
6 of which are not present in the photon vertex case. The  eight diagrams
that have a photon analogue
are shown in figure \ref{fig1}, and the  new ones are shown in figure
\ref{fig2}. One-loop
contributions
are formally of order
$\alpha$, but the magnitude of each diagram is in fact also
governed by the weak-boson or Higgs mass-factors like
 $m_\tau^2/M_{Z}^2$ or $m_\tau^2/M_{\Phi}^2$, so that the
Higgs contribution only modifies the real part of the result in less than a
1\%.  The main contributions come from the diagram with  $\nu WW$ and $W\nu\nu$ in
the loop. The final result is \cite{tau}:

\begin{equation}
a_\tau^w (M_Z^2)= - \;(2.10 + 0.61\, i) \times 10^{-6}
\label{wmdm}
\end{equation}

\noindent Notice the presence of an absorptive part of the same order as the
dispersive part
due to the fact that particles in the loop can be on-shell.

\mxfigura{7.25cm}{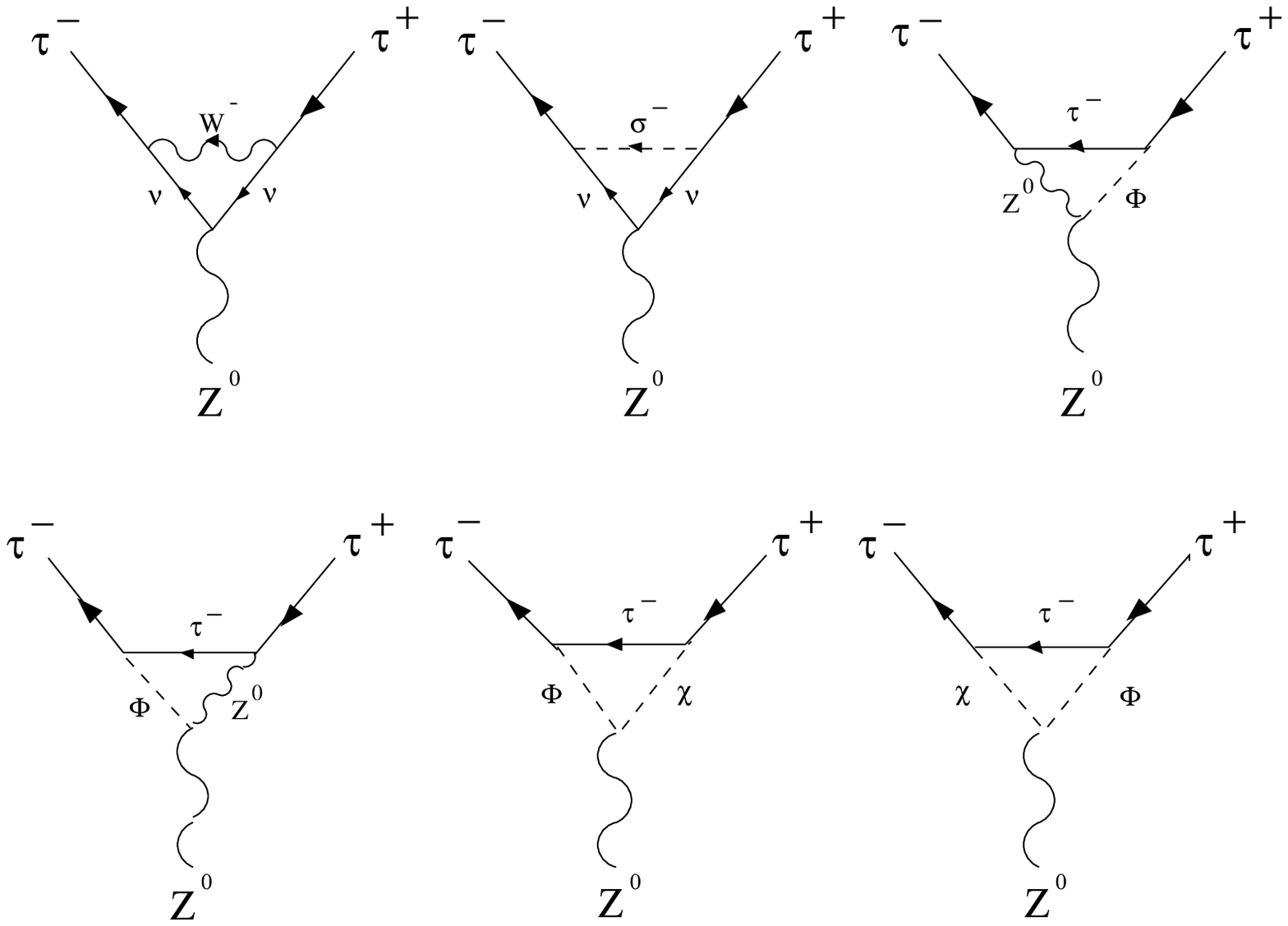}{Contributing Feynman diagrams to $a_\tau^w$ that
are not present for the AMDM (photon vertex).}{fig2}

On the other hand, from a very general argument, it was established
\cite{marciano} that for a fermion 
of mass $m_f$, contributions from new physics to 
the AWMDM at $\Lambda$-scale, must enter with a factor 
$\equiv m_f^2/\Lambda^2$, then it is clear that
precise experiments in the measurement of the AWMDM may provide  bounds for compositeness and also for 
the scale of new physics \cite{no,Illana1,Illana}.

For the (W)EDM, $d_\tau$, the theoretical predictions are much less precise.
In the SM the Kobayashi-Maskawa phase is the only source of $CP$-violation
and it is not possible to generate a non vanishing (W)EDM at one-loop
level; one has to go up to  three-loops \cite{wedm} to get a non vanishing 
contribution. In
extended models the situation changes and one can get a contribution to the
WEDM moment already at one-loop \cite{Illana2,concha,bern1}, so that a 
$CP$-violating signal coming from an appreciable (weak) electric dipole
moment will unambiguously lead to new physics. 

\section{SINGLE TAU POLARIZATION AT LEP1\hfill}

Using the lagrangean (\ref{lagrangean}), the tree level cross section for 
the process $e^+\, e^- \rightarrow \tau^+ \tau^-$ 
unpolarized scattering, at the $Z$-peak, can be written as:

\begin{equation}
\frac{d \sigma}{d  \Omega_{\tau^-}}=
\frac{d \sigma^{0}}{d  \Omega_{\tau^-}}
+\frac{d \sigma^{S}}
{d  \Omega_{\tau^-}}+\frac{d \sigma^{SS}}
{d  \Omega_{\tau^-}}
\label{cross1}
\end{equation}

\noindent where the first term collects the  spin independent terms, 

\begin{eqnarray}
&&\spa \frac{d \sigma^0}{d  \Omega_{\tau^-}}=
\frac{\alpha^2 \beta}{(4s_wc_w)^4}
\frac{1}{\Gamma_Z^2}\nonumber \\
&&\times \left[A_0+A_1\cos^2\theta_{\tau^-}+
A_2\cos\theta_{\tau^-}\right]\label{cross2}
\end{eqnarray}
with

\begin{eqnarray}
A_0&=&(v^2+a^2)\frac{2v^2+\beta^2(a^2-v^2)}{2s_w c_w}\label{a0}\\
A_1&=&(v^2+a^2)^2\beta^2\frac{1}{2s_w c_w}\label{a1}\\
A_2&=& 4a^2v^2\beta\frac{1}{s_w c_w}\label{a2}
\end{eqnarray}

\noindent The second one  takes into account the linear terms in  the 
spin \cite{tau,ttau},

\begin{eqnarray}
\frac{d \sigma^{S}}{d \Omega_{\tau^-}}&=&
\frac{\alpha^2\beta }{ 128 s_w^3 c_w^3}\frac{1}{\Gamma_Z^2}
\left[\  (s_-+s_+)_x X_+\right.\nonumber\\
&+&  (s_-+s_+)_y Y_++ 
  (s_-+s_+)_z Z_+  \nonumber\\
&+&\left.(s_--s_+)_y Y_-\right]
\label{cross3}
\end{eqnarray}

\noindent with 

\begin{eqnarray}
&&\hspace*{-.75cm} X_+= a\; \sin\theta_{\tau^-} \bigg\{\frac{-v \left[2v^2
+(v^2+a^2)\beta\cos
\theta_{\tau^-}\right]}{\gamma s_wc_w}\nonumber\\  
&&\hspace*{.75cm}+2\gamma\left[2v^2(2-\beta^2)\right.\nonumber\\
&&\hspace*{.75cm}\left. +(v^2+a^2)\beta\cos
\theta_{\tau^-}\right] \Re(a^w_{\tau})\bigg\}\label{x}\\
&&\spa Y_+=\;-2 v \gamma \beta \sin\theta_{\tau^-} 
\;[2 a^2\nonumber\\
&&\hspace*{.75cm}+(v^2+a^2) \beta \cos \theta_{\tau^-}]\;
 \Im(a_{\tau}^w) \label{y+}\\
&&\spa Y_-=2a\gamma\beta\sin\theta_{\tau^-} \left[
2v^2\right.\nonumber\\
&&\hspace*{.75cm}\left.+(v^2+a^2)\beta\cos\theta_{\tau^-}\right] 
(2 m_{\tau} d^w_{\tau}/e) \label{y-}\\
&&\spa Z_+=-\frac{va}{s_wc_w}
\left[(v^2+a^2)\beta(1+\cos^2\theta_{\tau^-})\right.\nonumber\\
&&\left.+2(v^2+\beta^2a^2)\cos\theta_{\tau^-}\ \right]
+2a\left[4v^2\cos\theta_{\tau^-}\right.\nonumber\\
&&\left. +(v^2+a^2)\beta(1+\cos^2\theta_{\tau^-})\right]
\Re(a^w_{\tau})\label{z}
\end{eqnarray}

\noindent where $\alpha$ is the fine structure constant, $\Gamma_Z$ is
the $Z$-width and $\gamma=M_Z/(2 m_\tau)$,
$\beta=(1-1/\gamma^2)^{1/2}$
are the dilation factor and $\tau$ velocity, respectively.
$s_\pm$ are the polarization vectors of $\tau^\pm$
in the proper reference frame and  $v=-1/2+2s^2_w$ and $a=-1/2$ are the
SM vector and axial vector $Z\, \tau^-\tau^+$ couplings.
We have neglected terms proportional to
the electron mass and kept up to linear terms in the weak dipole moments.
The reference frame is chosen  such that
the outgoing $\tau^-$ momenta is along
the $z$ axis and the incoming $e^-$ momenta is in the $x-z$ plane,
with $\theta_{\tau^-}$ the angle determined by these two momenta. Terms
with $(s_--s_+)_{x,y,z}$ factors in \eq{cross3}
 carry all the information
about the
$CP$-violating pieces of the lagrangean. Normal and transverse $\tau$
polarizations are zero at tree level in the SM, in the zero mass limit, due to
helicity arguments.

The spin-spin term, $\frac{d \sigma^{SS}}{d  \Omega_{\tau^-}}$, of 
\eq{cross1}, 
is not relevant in this calculation due to the fact that we are going to 
consider polarization asymmetries of a single $tau$ only. By summing up over the
polarization states of the other $tau$, it results in erasing this contribution. 

\subsection{Normal polarization ($P_N$)}

The normal polarization (along $y$-axis) of a single $tau$
($Y_\pm$ terms of \eq{cross3}) is even under parity. Then,
considering the transformation properties of the dipole moments described in
section \ref{properties}, only
$a\cdot v^2 \cdot d_{\tau}^w$
or $a^3 \cdot d_{\tau}^w$ (no $v$ {\it suppression}, in this case) terms
are allowed in $Y_-$ (see \eq{y-}),
in contrast to the case in the spin-spin correlation observables, where the
leading term is  $a^2 \cdot v \cdot d_{\tau}^w$.
The $Y_+$ term is $CP$-conserving and time reversal-odd; it is an
observable generated by a $T$-odd absorptive part of the magnetic 
moment,$\Im(a_{\tau}^w)$). The
dependence with $a^2 \cdot v \cdot \Im(a_{\tau}^w)$ or
$v^3 \cdot \Im (a_{\tau}^w)$ is associated with the fact that the normal
polarization is even under parity. 

\subsection{Transverse polarization ($P_T$)}

The transverse polarization  (along the
$x$-axis) of a single
$\tau$ ($X_+$ term of \eq{cross3}) is parity-odd
and time reversal-even. It can only arise from the
interference of both helicity conserving and
helicity flipping amplitudes.
The first term of $X_+$ in \eq{x} comes from
helicity flipping suppressed
($1/\gamma\equiv 2m_\tau/M_Z$)
 amplitudes in the Standard Model and the second one comes from
the $\gamma$-enhanced  chirality
flipping weak-magnetic tensorial $a_{\tau}^w$ vertex. Both contributions must
be proportional to an odd number of axial-vector couplings $a$
($a\, v^3$ or $a^3$). 

If one allows the WEDM $d_\tau^w$ to have an (absorptive) imaginary part, then
there is also a term (let us say $X_-$) proportional to this $T(+)$ imaginary
part \cite{Illana}. We are not going to take into account this contribution
because sizeable contributions coming from new physics at a high $\Lambda$
scale can not give any absorptive part at the $M_Z$ scale, so that such terms
must be obtained at higher orders and, in principle, must be much smaller.

\section{AZIMUTHAL ASYMMETRIES}

At LEP1 $\tau$ pairs decay before reaching the detectors and the energies
and momenta of their hadronic decay products can be measured.
In channels where both $\tau$'s
decay semileptonically, the $\tau$ direction can only be
reconstructed  up to a two fold ambiguity \cite{bernreuther} if
no high precision
measurement of both charged hadron tracks is made. It is this
ambiguity that
destroys the information coming from polarization
when looking
at the decay products. However, with the help of
micro-vertex detectors, a high resolution reconstruction
of these hadron-tracks is possible, then the $\tau$ direction
can be completely reconstructed \cite{kuhn}.  This opens new possibilities
to measure
the transverse and normal component of the
polarization from the angular distribution of single $\tau$
decay products. Therefore
we will only  consider semileptonic decay channels for both
$taus$.

From \eq{cross1} and \eq{cross3},  and following standard 
procedures \cite{tsai}, it is 
straightforward to get the expression for the  $e^+ e^- \rightarrow 
\tau^+\  \tau^-\rightarrow h^+_1 X\; h^-_2 \nu_\tau $ and 
$h^+_1 \bar{\nu_\tau}\; h^-_2 X$ cross
sections:

\begin{eqnarray}
&&\spa\frac{d\sigma(e^+e^-\rightarrow \tau^+\tau^-
\rightarrow h^+_1 X h^-_2\nu_\tau)}
{d(\cos\theta_{\tau^-})\, d\phi_{h^-_2}}\nonumber\\
&&=Br(\tau^-\rightarrow h^-_2\nu_\tau)Br(\tau^+
\rightarrow h^+_1X)\nonumber\\
&&\times \Bigg[
4 \frac{d\sigma^{0}}{d\Omega_{\tau^-}}+
\frac{\alpha^2\beta \pi}{128s_w^3 c_w^3\Gamma_Z^2}
\alpha_{h_2^-} (X_+\cos\phi_{h_2^-}\nonumber\\
&&+(Y_-+Y_+)\sin\phi_{h_2^-})\Bigg]\label{difcross1}\\
&&\spa \frac{d\sigma(e^+e^-\rightarrow \tau^+\tau^-
\rightarrow h^+_1\bar{\nu_\tau}h^-_2 X)}
{d(\cos\theta_{\tau^-})\, d\phi_{h^+_1}}\nonumber\\
&&=Br(\tau^-\rightarrow h^-_2
X)Br(\tau^+
\rightarrow h^+_1\bar{\nu_\tau})\nonumber\\
&&\spa\times\Bigg[
4 \frac{d\sigma^{0}}{d\Omega_{\tau^-}}+
\frac{\alpha^2\beta\pi }{128s_w^3 c_w^3\Gamma_Z^2}
\alpha_{h_1^+}(-X_+\cos\phi_{h_1^+}\nonumber\\
&&+(Y_--Y_+)\sin\phi_{h_1^+})\Bigg]\label{difcross2}
\end{eqnarray}

\noindent where the angle $\phi_h$ is the azimuthal hadron angle 
in the frame we have already defined.
All other angles have been 
integrated out. 
The longitudinal polarization term ($Z_+$) disappears 
when the polar angle $\theta_h$ of the hadron is 
integrated out.
For $\pi$ and $\rho$ mesons the magnitude of the parameter 
$\alpha_h$ is  $\alpha_\pi=0.97$ and  $\alpha_\rho=0.46$.

\subsection{Observables for the AWMDM}

With the $\tau$ direction fully reconstructed
in semileptonic decays, we can get information about the 
real part of the AWMDM, by
defining  the following asymmetry of the 
${\tau}$-decay products \cite{tau,ttau}:

\begin{equation}
A_{cc}^\mp=\frac{\sigma^\mp_{cc}(+)-\sigma^\mp_{cc}(-)}
{\sigma^\mp_{cc}(+)+\sigma^\mp_{cc}(-)}
\end{equation}

\noindent where

\begin{eqnarray}
&&\spa\sigma^{\mp}_{cc}(+)=
  \left(\int_{0}^{1} 
d(\cos \theta_{\tau^-}) 
\int_{-\pi/2}^{\pi/2}
 d \phi_{h^{\mp}}+\right.\label{s1cc}
\\
&&\spa\left.\int_{-1}^{0} d (\cos\theta_{\tau^-})
\int_{\pi/2}^{3\pi/2}d 
\phi_{h^{\mp}}\right) 
\frac{d \sigma}{d (\cos\theta_{\tau^-}) \; d \phi_{h^{\mp}}}\nonumber\\
&&\spa\sigma^{\mp}_{cc}(-)=
  \left(\int_{0}^{1} 
d(\cos \theta_{\tau^-}) 
\int_{\pi/2}^{3\pi/2}
 d \phi_{h^{\mp}}+\right.\label{s2cc}\\
&&\spa\left.\int_{-1}^{0} d (\cos\theta_{\tau^-})
\int_{-\pi/2}^{\pi/2}d 
\phi_{h^{\mp}}\right) 
\frac{d \sigma}{d (\cos\theta_{\tau^-}) \; d \phi_{h^{\mp}}}\nonumber
\end{eqnarray}

\noindent This asymmetry selects the $\cos\theta_{\tau^-}\cos \phi_{h^\mp}$ 
term of the cross section given in \eq{difcross1} 
and \eq{difcross2}, 
which is the leading one in the anomalous weak-magnetic moment $a_\tau^w$ 
(it comes with the couplings $a^3$),

\begin{eqnarray}
&&\spa A_{cc}^\mp= \mp \alpha_h \frac{s_wc_w(v^2+a^2)}{4 \beta a^3}
\nonumber\\
&&\hspace*{1cm}\times\left[  
\frac{-v}{\gamma s_wc_w} +2\gamma \; \Re(a^w_{\tau})\right]\label{acc}
\end{eqnarray}

\noindent Notice that it changes sign for $\tau^-$ and $\tau^+$.

Similarly, for the imaginary (absorptive) part of the AWMDM, one can define
an asymmetry that selects  the $\sin \phi_{h^\mp}$ term from $Y_+$ \cite{tau}:

\begin{eqnarray}
&& {A_s}^\mp= 
\frac{
\displaystyle{\int_{0}^{\pi} d \phi_h^{\mp}
\displaystyle{\frac{d\sigma}{d\phi_h^{\mp}}}}
-
\int_{\pi}^{2 \pi} d \phi_h^{\mp}
\displaystyle{\frac{d\sigma}{d\phi_h^{\mp}}}
}{\displaystyle{\int_{0}^{\pi} d \phi_h^{\mp}
\displaystyle{\frac{d\sigma}{d\phi_h^{\mp}}}}
+\int_{\pi}^{2 \pi} d \phi_h^{\mp}
\displaystyle{\frac{d\sigma}{d\phi_h^{\mp}}
}}\nonumber \\
&&= \mp \alpha_h \frac{3 \pi\gamma}{4} c_w s_w 
\frac{v}{a^2}\; \Im(a_{\tau}^w) 
\end{eqnarray}

For numerical results we  consider $10^7 Z$ events and
one $\tau$ decaying into
$\pi \; \nu_\tau$ or  $\rho \;  \nu_\tau$
({\it i.e.} $h_1 , \;h_2= \pi$ or $\rho$ in
(\ref{difcross1}) and (\ref{difcross2}) respectively),
while  summing up over the $\pi \; \nu_\tau$, $\rho \;
  \nu_\tau$ and
$a_1 \; \nu_\tau$  semileptonic  decay channels
of the $\tau$ for which the angular distribution is not observed (this amounts to
about 52\% of the total decay rate). Collecting events from the
decay of both $taus$, one gets a sensitivity (within 1 s.d.)\cite{tau}:

\begin{eqnarray} 
&&\spa  \abs{\Re(a_\tau^w)} \leq 4\cdot 10^{-4}\label{rawmdm}\\
&&\spa \abs{\Im(a_{\tau}^w)} \leq 1.1 \times 10^{-3}\label{iawmdm}
\end{eqnarray}

Comparing these values with  the SM predicted ones (\ref{wmdm}) it is clear 
that, if a large signal related to these observables is found, it should 
be attributed to physics beyond the SM.

\subsection{Observables for the WEDM}

The analysis of the
$tau$-decay products allows us to select the terms of the
cross sections (\ref{difcross1}) and (\ref{difcross2}) which carry the
relevant information about the $CP$-violating
effective coupling $d_\tau^w$. The leading term (the one with $a^3$)
is extracted by the asymmetry:

\begin{equation}
A_{sc}^\mp=\frac{\sigma^\mp_{sc}(+)-\sigma^\mp_{sc}(-)}
{\sigma^\mp_{sc}(+)+\sigma^\mp_{sc}(-)}
\label{asc}
\end{equation}

\noindent where $\sigma^\mp_{sc}(\pm)$ are defined similarly as in
\eqs{s1cc}{s2cc}
but changing the $\phi_{h^\mp}$ angular integration to:

\begin{eqnarray}
&&\spa\sigma^\mp_{sc}(+)=\sigma\left(\cos\theta_{\tau^-}\cdot
\sin\phi_{h^\mp}>0\right)\\
&&\spa\sigma^\mp_{sc}(-)=\sigma\left(\cos\theta_{\tau^-}\cdot
\sin\phi_{h^\mp}<0\right)
\end{eqnarray}
From \eqs{difcross1}{difcross2} we finally obtain:

\begin{equation}
A_{sc}^-=A_{sc}^+= \alpha_h \frac{\gamma}{2}  s_w c_w\frac{v^2+a^2}{a^3}
(2m_\tau d_\tau^w/e)\label{asc2}
\end{equation}

\noindent without any background from the Standard Model. It has the same
sign for both
$\tau^+$ and $\tau^¯$.

Under the same hypothesis as for the AWMDM one can get, from this asymmetry, 
the following bound to the WEDM \cite{ttau}:

\begin{equation}
\abs{d^w_\tau} \leq 2.3 \cdot 10^{-18} e \cdot cm
\end{equation}

The analysis made so far assumes there is no mixing among the weak dipole
moments in the defined asymmetries. Then, the bounds presented here
are the best one can get from azimuthal asymmetries. If one takes the
complete set of dipole moments in the calculation of the asymmetries one has
to either make a complete analysis with all the 
asymmetries \cite{L3} or construct
genuine $CP$-conserving (-violating) observables to disentangle the
different contributions. For example, a genuine $CP$-violating
observable is the asymmetry $A_{sc}$ as compared for the 
particle and its antiparticle 

\begin{equation}
A_{sc}^{CP}\equiv\frac{1}{2}(A_{sc}^-+A_{sc}^+)\label{truecp}
\end{equation}

\noindent What is tested
from the $A_{sc}^{CP}$-asymmetry is whether the normal polarizations
of both $taus$ are different. Within the contributions considered
in this paper,
they are opposite. This implies the equality of the decay-product
asymmetries (\ref{asc}), so $A_{sc}^{CP}=A_{sc}^+=A_{sc}^-$
and the observable is given only by the $CP$-violating term $d_\tau^w$,
eliminating the contribution from a ($v/a$) suppressed
$\Im(a_\tau^w)$ term, coming from the $Y_+$ sector of the normal polarization
(\ref{y+}). A similar $CP$-even observable can be obtained \cite{Illana} for
the $A_s^\mp$ asymmetry, $A_{s}=\left(A_s^--A_s^+\right)/2$,
which cancels the $CP$-odd (again $v/a$ suppressed) $d_\tau^w$ contribution
from the $Y_-$ sector (\ref{y-}) of $P_N$. 

\section{SPIN-SPIN CORRELATIONS}

Bounds on the $tau$ AWMDM  can be also obtained by
measuring spin correlation asymmetries
in the decay of a $Z$ to $\tau^+$ $\tau^-$. The $CP$-violating weak electric
dipole moment (WEDM) has been considered in Ref. \cite{bern2} by means of
momentum correlations of the decay products of the $\tau$ pair. 
For the $CP$-conserving
sector of the interaction described by lagrangean (\ref{lagrangean}), 
the spin-spin term of the
$e^+\, e^- \rightarrow \tau^+ \tau^-$ 
cross section,  at the $Z$-peak, can be written as:

\begin{eqnarray}
\frac{d \sigma^{SS}}{d \Omega_{\tau^-}}&=&
\frac{\alpha^2}{128 s_w^3 c_w^3}\frac{1}{\Gamma_Z^2} 
\left[\  (s_x^+s_x^-)\ {\cal XX }\right.\nonumber \\
&+&\left. (s_y^+s_y^-)\ {\cal YY}+(s_z^+s_z^-)\ {\cal ZZ}\right.\nonumber \\
&+&\left. (s_x^+s_y^-+s_y^+s_x^-)\ {\cal XY}\right.\nonumber\\
&+&\left. (s_x^+s_z^-+s_z^+s_x^-)\ {\cal ZX}\right.\nonumber \\ 
&+&\left. (s_y^+s_z^-+s_z^+s_y^-)\ {\cal ZY} \right]
\label{cross4}
\end{eqnarray}
where the coefficients $XX$, $XY$, $ZX$ \ldots carry all the information
about the Transverse-Transverse, Transverse-Normal, 
Longitudinal--Transverse \ldots spin correlations. 

Let us fix our attention on the spin correlation involving the 
transverse 
(within the production plane) and 
normal (to the production plane) components, relevant for the AWMDM:

\begin{eqnarray}
&&\spa {\cal XX} \equiv ({\cal XX})_0\, \sin^2\theta_{\tau^-}\\
&&\spa {\cal XY}\equiv ({\cal XY})_0\, \sin^2\theta_{\tau^-} \\
&&\spa {\cal ZX}\equiv ({\cal ZX})_0\sin\theta_{\tau^-}+({\cal
ZX})_1\sin2\theta_{\tau^-}\\
&&\spa {\cal ZY}\equiv ({\cal ZY})_0\sin\theta_{\tau^-}+({\cal
ZY})_1\sin2\theta_{\tau^-}
\end{eqnarray}

\noindent with

\begin{eqnarray}
&&\spa ({\cal XX})_0=(a^2+v^2)\nonumber \\
&& \times \left[\frac{\left(2v^2-\beta^2(v^2+a^2)\right)}
{2s_wc_w}-4v\Re(a_\tau^w)\right]\\
&&\spa ({\cal XY})_0=2a(a^2+v^2)\beta \Im(a_\tau^w)\label{cxy}\\
&&\spa  ({\cal ZX})_0= 2a^2v\beta\left[\frac{v}{s_wc_w\gamma}-
2\gamma\Re(a_\tau^w)\right]\nonumber
\end{eqnarray}
\begin{eqnarray}
&&\spa ({\cal ZX})_1= v(a^2+v^2)\nonumber \\
&& \times \left[\frac{v}{2s_wc_w\gamma}-
\gamma(2-\beta^2)\Re(a_\tau^w)\right]\\
&&\spa ({\cal ZY})_0= 4\gamma a v^2\beta^2\Im(a_\tau^w)\\
&&\spa ({\cal ZY})_1= \gamma\beta a(v^2+a^2)\Im(a_\tau^w)
\end{eqnarray}

These equations  show that
Transverse-Normal, $C_{TN}$  ($XY$ term),
and Longitudinal-Normal, $C_{LN}$  ($ZY$ terms),
spin correlations are directly proportional to the $T$-odd imaginary 
part of the
$tau$ weak magnetic dipole moment, while the  Longitudinal-Transverse  $C_{LT}$
 ($ZX$ terms) is proportional (except for a  small tree level contribution) to the
 real part of the dipole moment  (note that $\gamma \gg 1$).
 Transverse-Transverse $C_{TT}$ ($XX$ term) are also sensible to the real
part of the AWMDM but its contribution is not enhanced by the $\gamma$-factor. 

In particular the $XY$ spin correlation, $C_{TN}$, 
associated with the Transverse-Normal component of the $tau$ polarizations

\begin{equation}
<P_TP_N>=\frac{(XY)_0\sin^2\theta_\tau}{A_0+A_1\cos^2\theta_{\tau^-}+
A_2\cos\theta_{\tau^-}}\label{ctn1} 
\end{equation}

\noindent (with $A_0$, $A_1$ and $A_2$ defined in \eqs{a0}{a2}) is a
parity-odd and
time reversal-odd observable which, being generated by absorptive
parts of the  amplitude, must be proportional to the imaginary part of
the AWMM.  In the SM it also receives small contributions from the
interference of $\gamma$ exchange with the imaginary $Z$ exchange
amplitude \cite{nuria}.  When this contribution is subtracted from the
definition of $C_{TN}$, the measured value, from data collected in 1992-1994
running period,  by ALEPH \cite{Aleph} is: 

\begin{equation}
C_{TN}=-0.08\pm0.14(stat)\pm0.02(syst) 
\end{equation}

Using this data and 
the expression (\ref{ctn1}), in the $\beta\rightarrow 1$ limit, 

\begin{equation}
C_{TN}\approx \frac{(XY)_0}{A_0}=\frac{4 A}{(V^2+A^2)}\ c_w s_w 
\rm{Im}(a^w_\tau) 
\end{equation}

\noindent one gets the following bound on the imaginary part of the AWMM: 

\begin{equation}
|\rm{Im}(a^w_\tau)|<0.04 
\end{equation}
Up to now there is no measurement of the $C_{LT}$  and $C_{LN}$ 
correlations, that  are sensitive to the real  and imaginary part of 
the AWMDM. 

\section{CONCLUSIONS}

We have studied the physical content of the normal and
transverse  $\tau$
 polarizations for $\tau^+ \;\tau^-$ pairs produced from unpolarized
 $e^+\; e^-$
collisions at the $Z$-peak, and shown how their measurement
offers an opportunity to put bounds on the weak dipole moments induced
by models beyond the standard theory. For semileptonic decays, where
the $\tau$ direction can be reconstructed,
we have defined appropriate asymmetries in the azimuthal distribution of the
hadron from which one can measure the weak dipole moments. We have shown that
the best sensitivity one can expect in the measurement of these observables
is of the order of $10^{-4}$ ($10^{-3}$), for the real (imaginary) part of
the anomalous weak magnetic dipole moment, and $10^{-18}e\cdot cm$ for the weak
electric dipole moment. We have also shown an analysis of the spin-spin 
correlation terms that may also provide competitive independent bounds to 
the AWMDM. Nowadays sensitivities  are below those required in order
to measure the values predicted from the Standard Model.

\end{document}